\documentclass{llncs}
\usepackage{graphicx}

\usepackage{color}
\usepackage{tikz}
\usepackage{amsmath}
\PassOptionsToPackage{hyphens}{url}
\usepackage[hidelinks]{hyperref}

\usepackage{xspace}
\usepackage{units}

\usepackage{multirow}
\usepackage{booktabs}

\usepackage{appendix}

\usepackage[justification=centering]{subfig}

\begin{document}

\newcommand\copyrighttext{
       \footnotesize This is the author's version. The final authenticated publication is available online at\\ \url{https://doi.org/10.1007/978-3-030-72582-2_1}
}
\newcommand\copyrightnotice{
       \begin{tikzpicture}[remember picture,overlay]
              \node[anchor=south,yshift=50pt] at (current page.south) {\fbox{\parbox{\dimexpr\textwidth-\fboxsep-\fboxrule\relax}{\copyrighttext}}};
       \end{tikzpicture}
}

\title{Video Conferencing and Flow-Rate Fairness:\\A First Look at Zoom and the\\Impact of Flow-Queuing AQM}

\titlerunning{A First Look at Zoom and the Impact of Flow-Queuing AQM}

\author{Constantin Sander \and Ike Kunze \and Klaus Wehrle \and Jan Rüth}
\institute{Communication and Distributed Systems, RWTH Aachen University, Germany\\ \email{\{sander, kunze, wehrle, rueth\}@comsys.rwth-aachen.de}}

\maketitle

\newcommand{\sref}[1]{Section~\ref{sec:#1}}
\newcommand{\tab}[1]{Table~\ref{#1}}

\newcommand{\fig}[1]{Fig.~\ref{fig:#1}}
\newcommand{\fref}[1]{Fig.~\ref{fig:#1}}
\newcommand{\flabel}[1]{\label{fig:#1}}

\newcommand{\slabel}[1]{\label{sec:#1}}

\newcommand{\eref}[1]{Equation~\ref{sec:#1}}
\newcommand{\elabel}[1]{\label{sec:#1}}

\newcommand{\tref}[1]{Table~\ref{tab:#1}}
\newcommand{\tlabel}[1]{\label{tab:#1}}

\newcommand{\afblock}[1]{\noindent{\textbf{#1}}}
\newcommand{\takeaway}[1]{\noindent{\textbf{Takeaway.}} \textit{#1}}

\newcommand{\comments}[1]{\textbf{\color{red}{\{#1\}}}}
\newcommand{\constantin}[1]{\textbf{\color{red}{\{C: #1\}}}}
\newcommand{\todo}[1]{\textbf{\color{red}{\{TODO: #1\}}}}
\newcommand{\fix}[1]{{\color{red}{#1}}}
\newcommand{\ike}[1]{\textbf{\color{red}{\{I: #1\}}}}
\newcommand{\jan}[1]{\textbf{\color{red}{\{J: #1\}}}}
\newcommand{\klaus}[1]{\textbf{\color{red}{\{K: #1\}}}}

\newcommand{\mbps}{Mbps\xspace}

\newcommand{\zoomone}{ZC~1}
\newcommand{\zoomtwo}{ZC~2}

\copyrightnotice

\begin{abstract}
Congestion control is essential for the stability of the Internet and the corresponding algorithms are commonly evaluated for interoperability based on flow-rate fairness.
In contrast, video conferencing software such as Zoom uses custom congestion control algorithms whose fairness behavior is mostly unknown.
Aggravatingly, video conferencing has recently seen a drastic increase in use -- partly caused by the COVID-19 pandemic -- and could hence negatively affect how available Internet resources are shared.
In this paper, we thus investigate the flow-rate fairness of video conferencing congestion control at the example of Zoom and influences of deploying AQM.
We find that Zoom is slow to react to bandwidth changes and uses two to three times the bandwidth of TCP in low-bandwidth scenarios. %
Moreover, also when competing with delay aware congestion control such as BBR, we see high queuing delays.
AQM reduces these queuing delays and can equalize the bandwidth use when used with flow-queuing.
However, it then introduces high packet loss for Zoom, leaving the question how delay and loss affect Zoom's QoE.
We hence show a preliminary user study in the appendix which indicates that the QoE is at least not improved and should be studied further.

\end{abstract} %
\section{Introduction}
The stability of the Internet relies on distributed congestion control to avoid a systematic overload of the infrastructure and to share bandwidth.
Consequently, protocols that make up large shares of Internet traffic, such as TCP and QUIC, feature such congestion control mechanisms.

The COVID-19 pandemic and subsequent actions to limit its spread have now caused a drastic increase in traffic related to remote-working~\cite{feldmann:imc2020:lockdown}.
Of particular interest is the increasing share of video conferencing software which typically bases on UDP to conform to the inherent low-latency and real-time requirements which cannot be provided by TCP~\cite{brosh:ton10:delaytcprealtime,decicco:compnets:skypeinvestigation}.
Yet, UDP features no congestion control, meaning that the video conferencing software has to implement it on the application layer.
While this allows for adapting the video conference to the specific network conditions~\cite{garlucci:ton:googlewebrtc,decicco:compnets:skypeinvestigation}, such implementations can introduce unknown effects and undesired behavior when interacting with ``traditional'' congestion control.
Especially in light of the now increased share of the overall traffic, these tailored implementations can potentially pose a threat to Internet stability.

Thus, we investigate the interaction of real-world video conferencing software and traditional congestion control.
For our study, we choose Zoom as it has seen an enormous increase in traffic share by at least one order of magnitude from being marginally visible up to surpassing Skype and Microsoft Teams at certain vantage points~\cite{feldmann:imc2020:lockdown}. %
We focus on how Zoom reacts to loss and how it yields traffic to competing TCP-based applications.
We also study the impact of Active Queue Management (AQM) on the bandwidth sharing as it is of growing importance. %
Specifically, our work contributes the following:
\vspace{-0.2em}
\begin{itemize}
	\item We present a testbed-based measurement setup to study Zoom's flow-rate when competing against TCP CUBIC and BBRv1. %
	\item Comparing different bandwidths, delays, and queue sizes, we find that Zoom uses a high share on low-bandwidth links and that there are high queuing delays, even despite TCP congestion control trying to reduce it (e.g., BBR).
	\item We show that flow-queuing AQM reduces queuing delay and establishes flow-rate equality to a certain degree reducing Zoom's and increasing TCP's rate by dropping Zoom's packets, where the former is probably beneficial but the latter is probably detrimental for Zoom's QoE.
	Our preliminary user study shows that users do not see QoE improvements with flow-queuing AQM.
\end{itemize}
\vspace{-0.2em}
\afblock{Structure.}
Sec.~\ref{sec:bgrelwork} discusses the definition of fairness, as well as related work on general and video conferencing specific congestion control fairness analyses.
Sec.~\ref{sec:design} describes the testbed for our flow-rate equality measurements.
Sec.~\ref{sec:eval} shows our general results on Zoom and the impact of AQM on flow-rate equality, packet loss, and delay.
A preliminary user study evaluating the impact of AQM on the QoE can be found in the appendix.
Finally, Sec.~\ref{sec:conclusion} concludes this paper. %
\section{Background and Related Work}
\slabel{bgrelwork}
The interaction of congestion control algorithms, especially regarding fairness, is a frequent focus of research.
It has been thoroughly investigated for common TCP congestion control algorithms.
However, the definition of fairness itself has also been investigated and discussed.

\afblock{Fairness Definition.}
Most work relies on the conventional flow-rate definition of fairness: competing flows should get an equal share of the available bandwidth~\cite{jain:jainsfairnessindex}.
However, there are compelling arguments that flow-rate fairness is not an optimal metric~\cite{briscoe:ccr:fairness,ware:hotnets19:fairness} and new metrics such as harm~\cite{ware:hotnets19:fairness} propose to also consider the demands of applications and their flows.
We agree that flow-rate equality is no optimal metric for fairness as it ignores specific demands and the impact of delay, thus making it an outdated fairness estimate.

On the other hand, the notion of harm is hard to grasp as it requires (potentially wrong) demand estimates. %
Further, techniques such as AQM are demand unaware and flow-queuing even specifically aims at optimizing flow-rate equality, ignoring any actual application demands.
Hence, given the prevalence of flow-rate equality in related work and AQM techniques, we explicitly use flow-rate equality as our fairness metric to evaluate the precise impact of this metric on the application performance.
That is, we want to, e.g., see the impact on video conferencing when flow-queuing is used.
This naturally also means that results depicting an ``unfair'' flow-rate distribution are not necessarily bad.

\afblock{TCP Congestion Control.}
Many of the congestion control studies have especially looked at CUBIC~\cite{ha:sigops:cubic} and BBR~\cite{cardwell16:queue:bbr} and found that BBR dominates in under-buffered scenarios causing packet loss and making CUBIC back off, while it is disadvantaged in over-buffered scenarios~\cite{hock:icnp17:bbreval,kunze:tnsm:cpf,scholz:ifip18:deeperunderstandingbbr,ware:imc19:bbrmodel}.
Here, CUBIC, as a loss-based algorithm, fills the buffer and increases the queuing delay which makes BBR back off.
Introducing AQM, these behavior differences vanish.

\afblock{Impact of AQM.}
AQM mechanisms come with the potential of giving end-hosts earlier feedback on congestion, thus helping to reduce queuing delays, and there have been extended studies regarding their fairness (for a survey see~\cite{adamns:survandtut13:aqm}).
While some AQM algorithms are specifically designed to enable a fair bandwidth sharing (see~\cite{chatranon:globecom04:aqmfairness} for an overview and evaluation), generally, any AQM can be made to fairly share bandwidth with the help of fair queuing~\cite{demers:sigcomm89:fq}.
Today, this idea is most commonly implemented through a stochastic fair queuing (SFQ) which performs similar to a true fair queuing when the number of flows is limited.
In fact, several works (e.g., \cite{khademi:ietf88iccrg:aqm_fq,kunze:tnsm:cpf}) show that AQM using this SFQ (often called flow-queuing) can create flow-rate fairness while effectively limiting congestion, even though there are no comprehensive studies available in literature.

\subsection{Congestion Control For Video Conferencing} %
\label{sub:congestion_control_for_video_conferencing}
Loss-based congestion control, such as CUBIC, is not favorable to delay-sensitive real-time applications.
Hence, research has proposed several congestion control algorithms tailored to the needs of video conferencing.
However, in contrast to general-purpose congestion control, there is only limited research on its interaction mostly focusing on proposed algorithms with known intrinsics.

\afblock{Known Algorithms.}
For example, the Google Congestion Control (GCC)~\cite{garlucci:ton:googlewebrtc}, used in Google Chrome for WebRTC, was tested for flow-rate fairness~\cite{garlucci:ton:googlewebrtc,garlucci:ccr18:gccaqm}.
The results indicate that GCC shares bandwidth equally with CUBIC when using a tail-drop queue and also subject to the CoDel and PIE AQM algorithms.

There are similar findings for the Self-Clocked Rate Adaptation for Multimedia (SCReAM)~\cite{johannson:rfc8298:scream} congestion control algorithm.
It achieves an approximately equal share with a long-lived TCP flow on a tail-drop queue and yields bandwidth when using CoDel~\cite{johansson:rmcat:screamtest}.
Contrasting, the Network-Assisted Dynamic Adaptation (NADA) congestion control~\cite{zhu:ipvw13:nada} shares bandwidth equally when using a tail-drop queue, but uses bigger amounts when being governed by an AQM algorithm.

\afblock{Unknown Algorithms in Video Conferencing Software.}
However, many actually deployed real-world congestion control algorithms in video conferencing software are unknown and closed-source.
Thus, similar to our work, research also studies the externally visible behavior of video conferencing software.

De Cicco et al.~\cite{decicco:compnets:skypeinvestigation} investigate the behavior of Skype's congestion control and find that it is generally not TCP-friendly and claims more than its equal share.
Interestingly, Zhang et al.~\cite{zhang:infocom12:skypeprofiling} found that Skype yields a bigger share to competing TCP flows, but only after exceeding a certain loss threshold.
However, in contrast to work on TCP congestion control, these studies only consider limited scenarios and generally do not provide extensive evaluations (e.g., no AQM).

Other works focus even more only on aspects impacting the video conference, e.g., how the audio and video quality evolve subject to packet loss with unlimited rates~\cite{liotta:momm12:skypeqoe,xu:imc12:google-ichat-skype-inv} or very specific wireless settings~\cite{zhu:iwqos11:skypewimax}.

\takeaway{
Studies on general congestion control are not applicable to video conferencing.
Research on video conferencing software, on the other hand, mostly focuses on the concrete impact on its quality while the number of evaluation scenarios and the context to the general congestion control landscape is scarce.
}

We thus identify a need for a more thorough evaluation of real-world video conferencing congestion control that also considers the impact of different bandwidths, buffer sizes, or AQM on fairness.
For this purpose, we devise a methodology that centers around a configurable testbed which allows us to evaluate the behavior of the congestion control of Zoom.

\section{Measurement Design}
\slabel{design}
Research on congestion control fairness is often done using simulations or isolated testbeds to focus on the intrinsics of the algorithms.
In contrast, our work on Zoom forbids such an approach as the Zoom clients interact with a cloud-based backend that is responsible for distributing audio and video traffic.
Thus, to fully grasp the real-world performance of Zoom, we devise a testbed that connects to this backend while still letting Zoom's traffic compete with a TCP flow over a variety of network settings.
While we consequently have to take potential external effects into account, our testbed still allows us to control parameters, such as bottleneck bandwidth, queuing, and delay.

\subsection{Preliminaries}
For our investigations, we set up two Zoom clients which then connect to a joint Zoom conference via the Zoom backend running in a data center.
We find that free Zoom licenses use data centers operated by Oracle in the US, while our University license mostly connects to data centers operated by AWS in Europe.
We generally see that connections are established to at least two different AWS data centers, one in Frankfurt (Germany) and one in Dublin (Ireland).
As our upstream provider peers at DE-CIX in Frankfurt, we choose to focus on these connections to reduce the number of traversed links, thus minimizing the probability of external effects, such as changing routes or congestion.

\begin{figure}[t]
	\centering
\begin{tikzpicture}[every path/.append style={line width=0.1mm}, router/.pic={
	\fill[draw=black,fill=white] (0,0) circle (0.5);
	\draw[thin, shift={(0,0.4)},scale=0.15,rotate=-90] (0,0) -- (0.5,0.5) -- (0.5, 0.25) -- (1.5, 0.25) -- (1.5, -0.25) -- (0.5, -0.25) -- (0.5, -0.5) -- (0,0) -- (0.5,0.5);
	\draw[thin, shift={(-0.15,0)},scale=0.15,rotate=180] (0,0) -- (0.5,0.5) -- (0.5, 0.25) -- (1.5, 0.25) -- (1.5, -0.25) -- (0.5, -0.25) -- (0.5, -0.5) -- (0,0) -- (0.5,0.5);
	\draw[thin, shift={(0.15,0)},scale=0.15,rotate=0] (0,0) -- (0.5,0.5) -- (0.5, 0.25) -- (1.5, 0.25) -- (1.5, -0.25) -- (0.5, -0.25) -- (0.5, -0.5) -- (0,0) -- (0.5,0.5);
	\draw[thin, shift={(0,-0.4)},scale=0.15,rotate=90] (0,0) -- (0.5,0.5) -- (0.5, 0.25) -- (1.5, 0.25) -- (1.5, -0.25) -- (0.5, -0.25) -- (0.5, -0.5) -- (0,0) -- (0.5,0.5);
},
switch/.pic={\fill[draw=black,fill=white] (-0.5,-0.35) rectangle ++(1,0.7);
	\draw[thin, shift={(0.375,0.2)},scale=0.15,rotate=180] (0,0) -- (0.5,0.5) -- (0.5, 0.25) -- (2.5, 0.25) -- (2.5, -0.25) -- (0.5, -0.25) -- (0.5, -0.5) -- (0,0) -- (0.5,0.5);
	\draw[thin, shift={(-0.375,0.067)},scale=0.15,rotate=0] (0,0) -- (0.5,0.5) -- (0.5, 0.25) -- (2.5, 0.25) -- (2.5, -0.25) -- (0.5, -0.25) -- (0.5, -0.5) -- (0,0) -- (0.5,0.5);
	\draw[thin, shift={(0.375,-0.067)},scale=0.15,rotate=180] (0,0) -- (0.5,0.5) -- (0.5, 0.25) -- (2.5, 0.25) -- (2.5, -0.25) -- (0.5, -0.25) -- (0.5, -0.5) -- (0,0) -- (0.5,0.5);
	\draw[thin, shift={(-0.375,-0.2)},scale=0.15,rotate=0] (0,0) -- (0.5,0.5) -- (0.5, 0.25) -- (2.5, 0.25) -- (2.5, -0.25) -- (0.5, -0.25) -- (0.5, -0.5) -- (0,0) -- (0.5,0.5);
},
host/.pic={\fill[draw=black,fill=white] (-0.2,-0.4) rectangle ++(0.4,0.8);
	\draw[] (-0.15, -0.3) -- (0.15, -0.3);
	\draw[] (-0.15, -0.2) -- (0.15, -0.2);
	\draw[] (-0.15, -0.1) -- (0.15, -0.1);
	\draw[] (-0.15,  0.0) -- (0.15,  0.0);
	\fill   ( 0.1,  0.2) circle (0.05);
},
shost/.pic={\fill[draw=black,fill=black!30!] (-0.2,-0.4) rectangle ++(0.4,0.8);
	\draw[] (-0.15, -0.3) -- (0.15, -0.3);
	\draw[] (-0.15, -0.2) -- (0.15, -0.2);
	\draw[] (-0.15, -0.1) -- (0.15, -0.1);
	\draw[] (-0.15,  0.0) -- (0.15,  0.0);
	\fill   ( 0.1,  0.2) circle (0.05);
},
btln/.pic={\fill[draw=black,fill=white] (-1,-0.25) rectangle ++(2,0.5);},
cloud/.pic={
	\draw[line width=0.8pt,postaction={white,fill}]
		(-0.5,  0.0) circle (0.3)
		( 0.5,  0.0) circle (0.3)
		( 0.0,  0.3) circle (0.3)
		( 0.0,  0.0) ellipse (0.6 and 0.3)
		(-0.4,  0.15) circle (0.3)
		( 0.0,  -0.2) circle (0.3)
		( 0.4,  -0.2) circle (0.3)
		( 0.4,  0.2) circle (0.3)
		(-0.4,  -0.2) circle (0.3);
}]
\def\router(#1,#2){(#1,#2) pic {router}}
\def\switch(#1,#2){(#1,#2) pic[scale=0.75] {switch}}
\def\host(#1,#2){(#1,#2) pic[scale=0.75] {host}}
\def\shost(#1,#2){(#1,#2) pic {shost}}
\def\cloud(#1,#2){(#1,#2) pic[scale=0.9] {cloud}}
\def\btln(#1,#2){(#1,#2) pic {btln}}

\draw[ultra thick] \host(-3.0,0.5) node[anchor=east, outer sep=5, text width=2.1cm,align=right]{Zoom Client 1\\(\zoomone{})} -- (-2.0,0.0) -- \host(-3.0,-0.5) node[anchor=east, outer sep=5]{TCP Client};
\draw[ultra thick] \host(2.0,0.5) node[anchor=west, outer sep=5, text width=2.1cm,align=left]{Zoom Client 2\\(\zoomtwo{})} -- ( 1.0,0.0) -- \host( 2.0,-0.5) node[anchor=west, outer sep=5, text width=2.1cm,align=left]{\\TCP Server};

\draw[ultra thick] ( 1.0,0.0) -- \cloud(4.5,0.0) node {Internet};
\draw[opacity=0.0] \cloud(-5.75,0.0) node {Internet};

\draw[ultra thick] \switch(-2.0,0.0) -- \btln(-0.5,0.0) node{Bottleneck} -- \switch(1.0,0.0);

\end{tikzpicture}
 	\caption{Testbed setup representing a dumbbell topology}
	\flabel{testbed}
\end{figure}
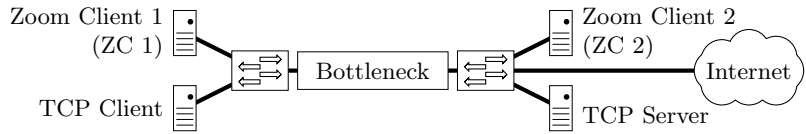

\subsection{Testbed Setup}
As shown in \fref{testbed}, our testbed uses a dumbbell topology and consists of five dedicated machines.
In the center, one machine serves as the configurable bottleneck link over which Zoom Client 1 (\zoomone{}) connects to the Zoom backend to join a conference with Zoom Client 2 (\zoomtwo{}).
Our two remaining machines (TCP Client, TCP Server) operate a concurrent TCP flow to assess competition.

\afblock{Testbed Interconnection.}
All our machines are interconnected using \unit[1]{Gbps} Ethernet links. %
The uplink to our university's network is \unit[10]{Gbps} which in turn connects to the German Research Network (DFN) via two \unit[100]{Gbps} links.
The DFN then peers at DE-CIX with, e.g., AWS.
We can thus be reasonably sure that our configurable bottleneck machine represents the overall bottleneck.

\afblock{Shaping the Bottleneck.}
We configure our bottleneck using Linux’s traffic control (TC) subsystem similar to~\cite{rueth:tma19:cpf} to create network settings with different bandwidths, delays, queue sizes, and queue management mechanisms.
For rate-limiting, we use token bucket filters with a bucket size of one MTU (to minimize bursts) on the egress queues in both directions.
Similarly, we also configure the AQM on the egress queues.
Delay is modeled on the ingress queues using intermediate function blocks (ifbs) and netem.
We first create an additional ingress qdisc via ifb and add the delay to the egress of this ifb via netem.
This technique is necessary as netem is not directly compatible with AQM qdiscs~\cite{bufferbloatnet:codelpractices} and usage of netem on the end-hosts would cause issues due to TCP small queues~\cite{cardwell:bbreval}.
Further, we add no artificial jitter, as this causes packet reorderings, as such, jitter is only introduced through the flows filling the queue itself.

\afblock{Balancing RTTs.}
Our testbed compensates for differing RTTs and ensures that the Zoom and the TCP flow have the same RTT, a requirement for the common flow-rate equality definition.
For this, we first measured the average delay between different AWS hosts and \zoomone{} as well as between TCP Client and TCP Server prior to our experiments.
We then adapted the netem delay accordingly such that the TCP flow and the flow between \zoomone{} and AWS have about the same RTT when the queue is empty.
By adapting the delay prior to our experiments, we avoid skewing the initial RTT of flows which we presume to be important for Zoom's congestion control, but accept a potential bias due to changing hosts at AWS which we cannot predict prior to establishing our video conferences.
However, the relative error of this bias should be insignificant as we emulate rather large artificial RTTs.

\subsection{Fairness Measurement Scenarios and Procedure}
With our measurements, we aim to represent video conferences from a low-speed residential access where Zoom's video flow and a TCP flow (e.g., a movie download) compete.
The used parameters are shown in \tref{params}.

\begin{table}[t]
	\centering
	\setlength{\tabcolsep}{0.38em}
	\begin{tabular}{ccccccc}
		\toprule
		BW\,[Mbps]    & RTT\,[ms]       & QSize\,[BDP] & AQM & CC & Order & Direction         \\
		\midrule
		\multirow{2}{*}{0.5, 1, 2, 4} & \multirow{2}{*}{30, 50} & \multirow{2}{*}{0.5, 2, 10} & Tail-Drop & CUBIC & Zoom first & Downlink \\
		& &    & (FQ\_)CoDel & BBRv1    & TCP first   & Uplink       \\
		\bottomrule
	\end{tabular}
	\vspace{0.5em} %
	\caption{Parameter configuration for our testbed}
	\vspace{-0.5em} %
	\tlabel{params}
\end{table}

The lowest bandwidth (\unit[0.5]{\mbps}) falls slightly below Zoom's requirements of \unit[0.6]{\mbps}~\cite{zoom:bandwidth}.
Yet, we argue that it also has to behave sound in out-of-spec cases.

We shape the bandwidth symmetrically, which is atypical for a residential connection, but study the up- and downlink separately.
We also adjust and balance the minimum RTT (min-RTT) symmetrically as described before.
As queue sizes, we use multiples of the BDP, i.e., 0.5, 2, and 10$\times$ the BDP.
When investigating AQM, we use 2$\times$BDP as AQM algorithms require headroom to operate, and adopt the TC Linux defaults for CoDel (target 5ms and interval 100ms).
Further, we vary which flow starts first to investigate late-comer effects.

\afblock{Overcoming transient states.}
For our measurements, we want to avoid transient phases.
As such, we usually wait in the order of half a minute after activating each flow to stabilize.
We then start a \unit[60]{s} measurement period in which we capture all exchanged packets, measure the queuing delay, and also observe the queue sizes at the bottleneck using a small eBPF program.

\afblock{Video Conference.}
The Zoom video conference itself is established between \zoomtwo{} and \zoomone{} (ensuring connectivity via AWS in Frankfurt).
As their video feeds, both clients simulate a webcam via \texttt{v4l2loopback}~\cite{v4l2loopback}.
To rule out effects of video compression on the congestion control behavior of Zoom, we ensure a constant video data rate by using uniform noise as our video input. %

Every scenario is repeated 30 times and retried where, e.g., Zoom restarts due to high loss.
The measurements were made from July 2020 to October 2020 on Linux \texttt{5.4.0-31} with Zoom version \texttt{5.0.408598.0517}.
To observe variations, we sort the points in the following scatterplots chronologically from left to right.

\afblock{Equality Metric.}
We measure flow-rate equality using the metric of our prior work~\cite{rueth:tma19:cpf}.
In contrast to, e.g., Jain's fairness index~\cite{jain:jainsfairnessindex}, this metric shows which flow over-utilizes the bottleneck by how much.
The metric is defined as:
\begin{equation*}
\textsf{flow-rate equality} =
\begin{cases}
1 - \frac{bytes(TCP)}{bytes(Zoom)}, & \text{if $bytes(Zoom) \geq bytes(TCP)$}\\
- 1 + \frac{bytes(Zoom)}{bytes(TCP)}, & \text{otherwise}
\end{cases}
\elabel{fairness}
\end{equation*}
$\textsf{flow-rate equality}$ lies in the interval of $[-1,1]$.
With $0$ both flows share the bandwidth equally, while $1$/$-1$ means that Zoom / TCP monopolizes the link.

Please note that flow-rate equality is likely \emph{not} the desired metric to depict a fair service enablement.
For example, Zoom simply needs a certain data-rate to deliver its service, as such flow-rate equality should likely not be used to establish fairness, e.g., in an AQM.
Nevertheless, we still opted for this metric to i) judge what happens when an AQM tries to utilize this metric, and ii) investigate the bandwidth demand and the ability of the congestion controller to seize the required bandwidth as well as the side effects in doing so.

\section{Zoom Inter-Protocol Fairness Results}
\slabel{eval}
In the following we present our findings on the behavior of Zoom by first analyzing its general congestion reaction (\sref{eval:general}).
We then discuss how \zoomone{} competes with a TCP flow in scenarios without AQM at low bandwidths subject to different queue sizes (\sref{eval:droptail}).
We further evaluate the effects of using CoDel (\sref{eval:codel}) and FQ\_CoDel (\sref{eval:fq_codel}) AQM.
Lastly, we show results of a small-scale user study that investigates the effects of FQ\_CoDel on the actual QoE, which can be found in the appendix to this work (Appendix~\ref{sec:userstudy}).

Before conducting our experiments, we first verify the standalone throughput of TCP and Zoom in our scenarios.
We find that TCP achieves a utilization above 80\% in almost all cases except for 3 outliers out of 4800 runs.
Similarly, Zoom's throughput for the AQM scenarios only changes by at most 10\%.
The following differences in flow-rate equality are thus mainly due to the interaction of the congestion control algorithms and not rooted in our settings.

\subsection{General Observations on Zoom's Behavior}
\label{sec:eval:general}
\begin{figure}[t]
	\centering
	\includegraphics{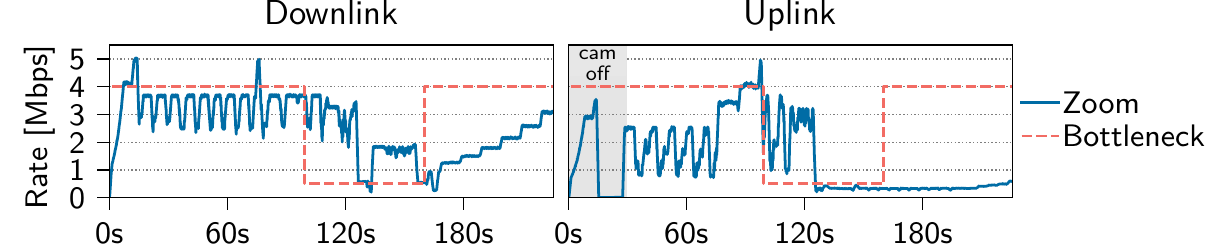}
	\caption{Zoom video flow behavior for a \unit[50]{ms} RTT and a 10$\times$BDP tail-drop queue. Bandwidth (orange) is varied from \unit[4]{\mbps} to \unit[0.5]{\mbps} and back to \unit[4]{\mbps}.}
	\vspace{-1em}
	\flabel{behavior}
\end{figure}
We first observe the behavior of a single Zoom flow without competition in a scenario with a \unit[50]{ms} RTT and a 10$\times$BDP tail-drop queue.
\fref{behavior} shows Zoom's video send rate when varying the bandwidth (orange) from \unit[4]{\mbps} to \unit[0.5]{\mbps} and back.
At first, Zoom's backend (left) sends at slightly less than \unit[4]{\mbps} while the Zoom client (right) sends at $\sim$\unit[2.5]{\mbps}. %
In both cases, the queue is empty.
Similar to BBR~\cite{cardwell16:queue:bbr}, Zoom seems to repeatedly probe i) the bandwidth by increasing its rate and ii) the min-RTT by reducing its rate.

Once we reduce the bandwidth to \unit[0.5]{\mbps}, both Zoom entities keep sending at $\sim$\unit[3.5]{\mbps}, thus losing many packets and filling the queue.
After $\sim$30s, Zoom reduces its rate to \unit[0.5]{\mbps}.
Surprisingly, the backend again increases the rate by a factor of 4 shortly thereafter.
After resetting the bandwidth to \unit[4]{\mbps}, Zoom slowly increases its rate on the uplink and faster on the downlink.

Packet loss and increased queuing delays do not seem to directly influence Zoom's sending behavior.
However, Zoom occasionally restarted the video conference completely, stopping sending and reconnecting to the backend with a new bandwidth estimate not overshooting the bottleneck link.
We filtered these occurrences from the following results as the time of reconnecting would influence our metric and also the meaning of our ``Zoom first'' scenario.

We also changed the min-RTT from 50ms to 500ms instead of the bandwidth.
We did not see any obvious reaction, although we expected that Zoom backs off to wait for now delayed signaling information or to reduce potential queuing.

To summarize, Zoom handles up- and downlink differently and does not seem to directly react on increased queuing or loss, instead reacting slowly which leads to big spikes of loss. %
We next investigate how this impacts competing flows.

\subsection{Competition at Tail-Drop Queues}
\label{sec:eval:droptail}
\afblock{Undersized Tail-Drop Queue.}
We first examine Zoom's behavior when competing at a 0.5$\times$BDP tail-drop queue against TCP CUBIC and BBR.
The scatterplots in \fig{eval:fairness_pfifo_bdp5} show our \textsf{flow-rate equality} %
 for downlink (a) and uplink (b).

\begin{figure}[t]
	\centering
	\subfloat[\flabel{eval:fairness_pfifo_bdp5:downlink}Results for competition on the downlink]{\includegraphics{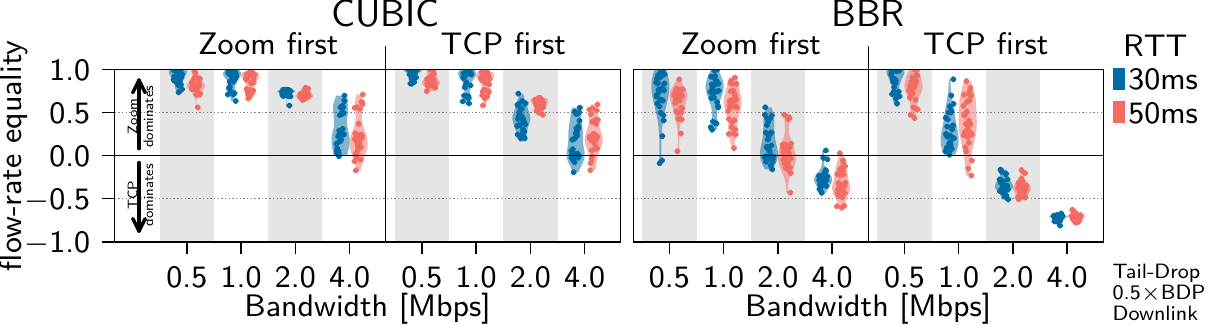}}

	\subfloat[\flabel{eval:fairness_pfifo_bdp5:uplink}Results for competition on the uplink]{\includegraphics{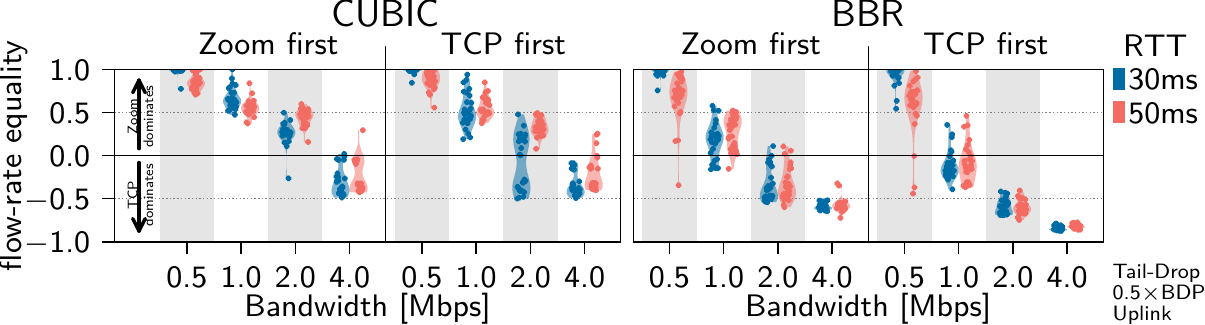}}
	\caption{Flow-rate equality for Zoom competing at a 0.5 $\times$ BDP queue with TCP.}
	\vspace{-1em}
	\flabel{eval:fairness_pfifo_bdp5}
\end{figure}

\textbf{Downlink.}
Zoom uses a disproportionate bandwidth share on the downlink with bottleneck bandwidths $\leq$ \unit[1]{\mbps}.
The flow-rate equality is mostly above 0.5, i.e., Zoom's rate is more than twice the rate of the TCP flow.
For higher bandwidths, Zoom yields more bandwidth.
Additionally, we can see that TCP flows starting first result in slightly better flow-rate equality.
For CUBIC, equality values of around 0 can be first seen at \unit[4]{\mbps}.
For BBR, equality values of around 0 can already be seen at \unit[2]{\mbps}.
However, when being started first and at \unit[4]{\mbps}, BBR disadvantages Zoom significantly.

\textbf{Uplink.} For the uplink, the equality values are comparable, but in total lower.
This means that the TCP flows claim more bandwidth (especially with BBR) and Zoom seems to act less aggressive.
We posit that Zoom's congestion control might be adapted to the asymmetric nature of residential access links.

The queuing delays on the down- and uplink mostly exceed 50\% of the maximum (not shown).
We attribute this to the TCP flows as i) CUBIC always fills queues, and ii) BBR overestimates the available bandwidth when competing with other flows~\cite{ware:imc19:bbrmodel} and then also fills the queue plus iii) Zoom reacting slowly.

\afblock{Slightly Oversized Tail-Drop Queues.}
When increasing the buffer size to 2$\times$BDP, the results are surprisingly similar (and thus not visualized).
CUBIC can gather a slightly larger bandwidth share, which we attribute to its queue-filling behavior.
However, Zoom still holds twice the bandwidth of the TCP flows at links with $\leq$\unit[1]{\mbps}, i.e. the equality values mostly exceed 0.5.
Only on faster links, CUBIC can gain an equal or higher bandwidth share.
For BBR, equality values are closer to 0 for bandwidths below \unit[2]{\mbps}, i.e., Zoom as well as BBR dominate less.
For higher bandwidths, the results are equivalent to before.
Also the avg. queuing delay rises to about 75\% due to filled queues as before.

\afblock{Overlarge Tail-Drop Queues.}
Next, we study the flow-rates for large queues of 10$\times$BDP.
\fig{eval:fairness_pfifo_bdp100} shows the results for downlink (a) and uplink (b).
\begin{figure}[t]
	\centering
	\subfloat[\flabel{eval:fairness_pfifo_bdp100:downlink}Results for competition on the downlink]{\includegraphics{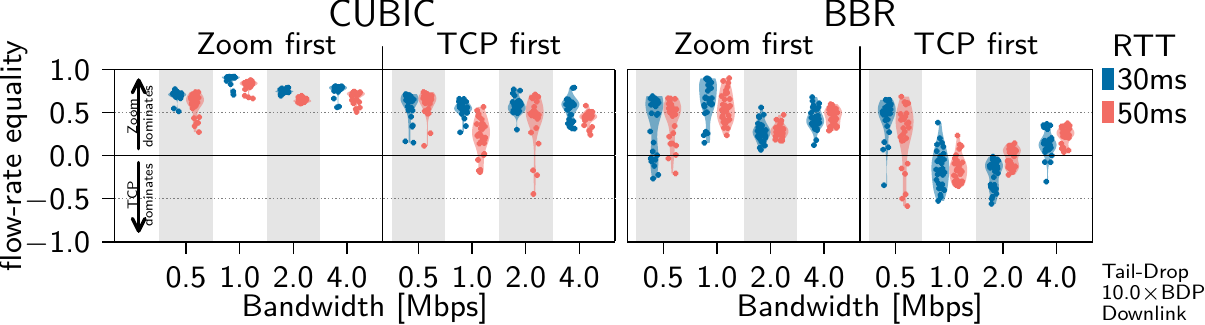}}

	\subfloat[\flabel{eval:fairness_pfifo_bdp100:uplink}Results for competition on the uplink]{\includegraphics{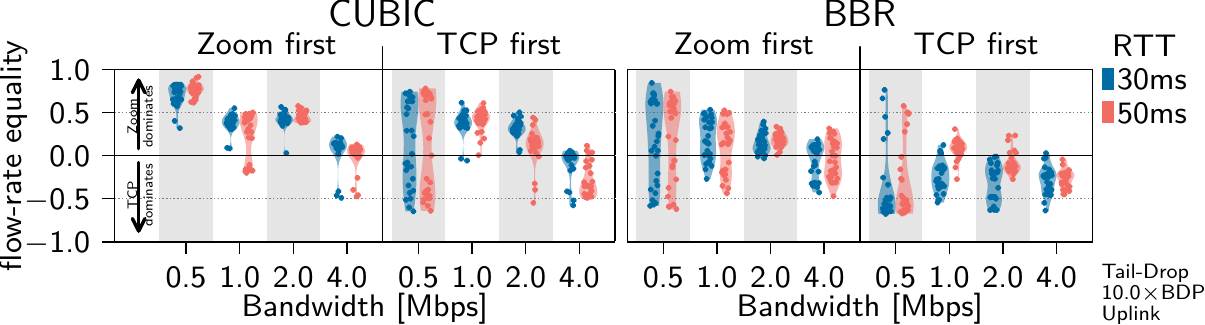}}
	\caption{Flow-rate equality for Zoom competing at a 10 $\times$ BDP queue with TCP.}
	\flabel{eval:fairness_pfifo_bdp100}
\end{figure}

\textbf{Downlink.}
Contrary to our expectation, there is no significant improvement in flow-rate equality for the downlink.
Zoom still uses a high bandwidth share and CUBIC's queue-filling behavior does not result in a larger share.
Compared to the previous scenarios, the equality values are not decreasing significantly when Zoom starts first and it even uses more bandwidth than before for the \unit[4]{\mbps} setting.
For TCP CUBIC starting first, equality values now spread around 0.5, regardless of the bandwidth.
For Zoom starting first, BBR barely reaches values below zero.

\textbf{Uplink.}
The scenario looks completely different for the uplink.
Zoom yields bigger parts of the bandwidth to CUBIC and even reduces on one third of the bandwidth when BBR starts first.
This is surprising, as BBR is known to be disadvantaged in this overbuffered scenario~\cite{hock:icnp17:bbreval}.
We also checked if changes between the BBR code used in~\cite{hock:icnp17:bbreval} and our Linux Kernel 5.4 could explain this difference, but the basic principle \emph{seems} to be unaltered.
Still, we remark that the BBR codebase has seen significant changes since~\cite{hock:icnp17:bbreval} and we are not aware of any investigations how these changes affect BBR's properties.

\begin{figure}[t]
	\centering
	\includegraphics{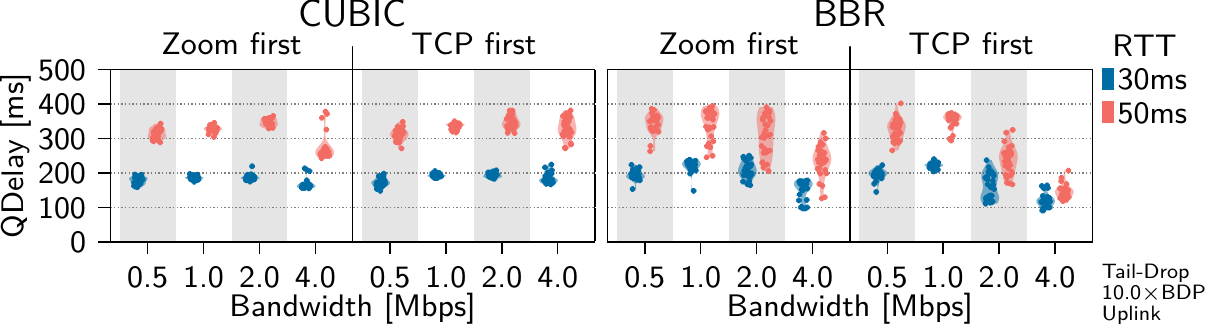}
	\caption{Queuing delay for Zoom competing at a 10 $\times$ BDP queue on the uplink.}
	\flabel{eval:delay_pfifo_bdp100}
\end{figure}

The queuing delay, shown in \fref{eval:delay_pfifo_bdp100} for the uplink, still reaches about 75\% of the maximum queuing delay for CUBIC and BBR in low-bandwidth scenarios where delay is slightly smaller on the uplink than on the downlink.
BBR seems to be able to reduce queuing delay in the higher bandwidth region, but we expected that BBR would reduce the queuing delay more strongly in all scenarios.

\takeaway{We can see that Zoom is unfair w.r.t. flow-rate to CUBIC in low-bandwidth scenarios with \unit[1.0]{\mbps} and less, although Zoom is less aggressive on the uplink.
As BBR is more aggressive, it gains higher rates in these situations -- also on the downlink.
However, all scenarios have in common that the queuing delay is significantly increased being detrimental to video conferencing.}

\begin{figure}[t]
	\centering
	\includegraphics{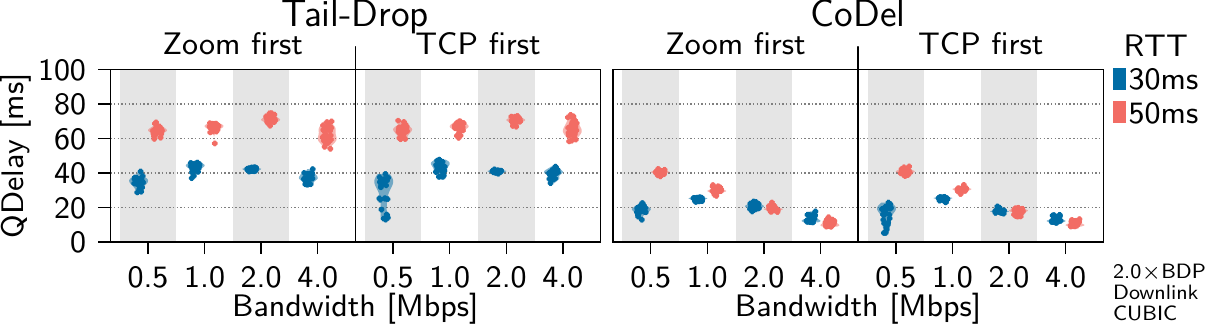}
	\caption{Queuing delay for Zoom+CUBIC competing at a tail-drop/CoDel queue.}
	\vspace{-1em}
	\flabel{eval:codel}
\end{figure}

\subsection{Competition at CoDel Queues}
\label{sec:eval:codel}
Using AQM might be beneficial, given the increased queuing delays.
Hence, we study Zoom and TCP flows competing at CoDel queues.
We expect significant changes in flow-rate equality as CoDel drops packets early to signal congestion.

Yet, our results are very similar to the 2$\times$BDP tail-drop queue, thus we do not show them here.
They only slightly shift towards CUBIC.
However, CoDel keeps its promise of reduced queuing delays, as shown in \fig{eval:codel}:
The queuing delay of Zoom competing with CUBIC (BBR looks similar) at 2$\times$BDP queues roughly halves when CoDel is used at \unit[0.5]{\mbps}.
For higher bandwidths, the effect is even stronger.
This is potentially beneficial for real-time applications.

\takeaway{
All in all, CoDel does not significantly alter the flow-rate distribution.
However, it keeps its promise of reducing the experienced queuing delays.
}

\begin{figure}[t!]
	\centering
	\subfloat[\flabel{eval:fq_codel:downlink}Results for competition on the downlink]{\includegraphics{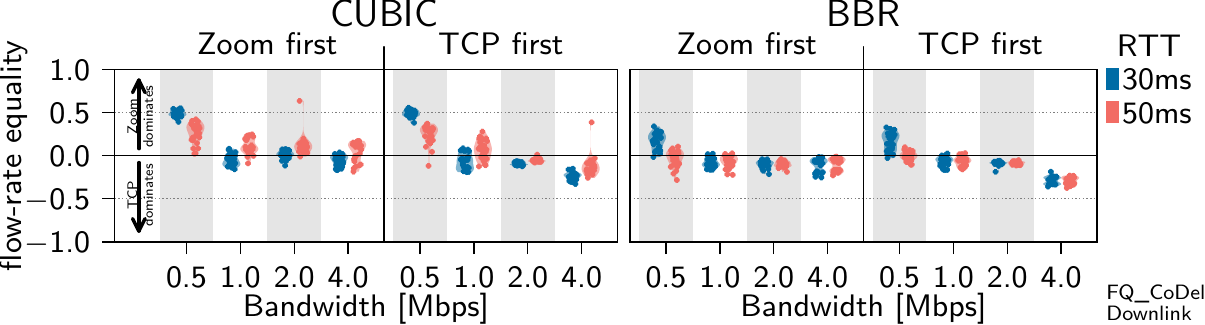}}

	\subfloat[\flabel{eval:fq_codel:uplink}Results for competition on the uplink]{\includegraphics{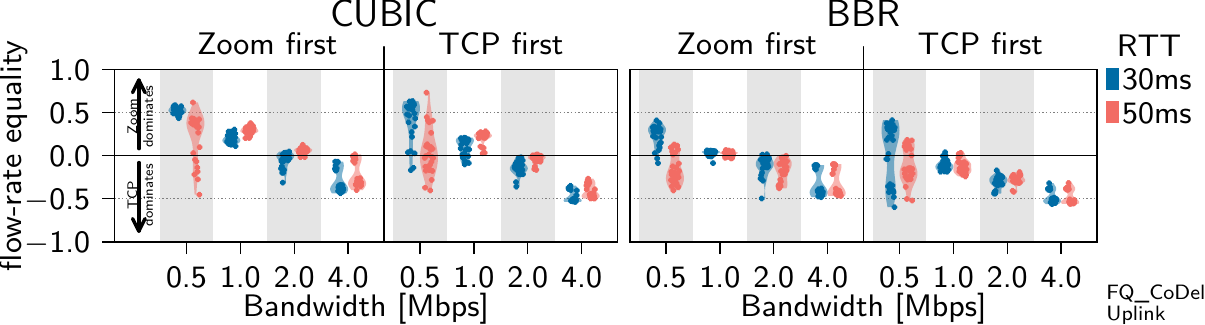}}
	\caption{Flow-rate equality for Zoom competing with TCP at an FQ\_CoDel queue}
	\vspace{-1em}
	\flabel{eval:fq_codel}
\end{figure}

\subsection{Competition at FQ\_CoDel Queues}
\label{sec:eval:fq_codel}
To enforce flow-rate equality, we next apply FQ\_CoDel to the queue.
FQ\_CoDel adds stochastic fair-queueing to CoDel, i.e., it isolates flows into subqueues, applies CoDel individually, and then serves the queues in a fair manner.

While the queuing delays are equivalent to CoDel and thus not shown, our flow-rate equality metric significantly shifts towards TCP in most conditions as shown in \fig{eval:fq_codel} for uplink (a) and downlink (b).
For example, the downlink results mostly range from 0.3 to -0.3 compared to prior findings of Zoom dominating.
The biggest advance for Zoom remains in the \unit[0.5]{\mbps} setting.

On the uplink, equality differs.
Zoom yields bandwidth when using BBR in mostly all cases except for bandwidths $\leq$ \unit[1.0]{\mbps}.
For CUBIC, also no perfect equalization can be seen.
For bandwidths above \unit[2.0]{\mbps} CUBIC gets bigger shares, below this threshold, vice versa.
We deduct this to Zoom being more careful on the uplink and not using the whole probed bandwidth, leaving a gap.

\begin{table}[t]
	\centering
	\setlength{\tabcolsep}{0.5em}
	\begin{tabular}{rcccccc}
		\toprule
		& \multicolumn{2}{c}{Tail-Drop}  & \multicolumn{2}{c}{CoDel} & \multicolumn{2}{c}{FQ\_CoDel} \\
		\cmidrule(lr){2-3}\cmidrule(lr){4-5}\cmidrule(lr){6-7}
		& Dropped  & Received & Dropped & Received & Dropped  & Received  \\
		\midrule
		TCP CUBIC & 188.0 & 816.5  & 190.0 & 935.5   & 250.5 & 1260.5 \\
		Zoom      & 331.0 & 2824.0 & 515.5 & 2852.5  & 903.5 & 2880.5 \\
		\bottomrule
	\end{tabular}
	\vspace{0.5em} %
	\caption{Median number of packets received and dropped for CUBIC and Zoom at a \unit[0.5]{\mbps}, \unit[50]{ms}, 2$\times$BDP bottleneck on the downlink (Zoom started first).}
	\vspace{-0.5em} %
	\tlabel{drops}
\end{table}

\afblock{Zoom's Reaction to Targeted Throttling.}
As we could see, FQ\_CoDel allows to share bandwidth between Zoom and competing TCP flows after a bottleneck more equally.
However, it is unclear whether Zoom reduces its rate or whether the AQM is persistently dropping packets, specifically in the low-bandwidth scenarios.
We hence show the dropped and sent packets for CUBIC and Zoom over 60s in \tref{drops} for the \unit[0.5]{\mbps} bottleneck with 2$\times$BDP queue and \unit[50]{ms} RTT.
We can see that Zoom does not reduce its packet-rate from a tail-drop queue up to FQ\_CoDel.
Instead, the AQM drops packets increasingly.

\takeaway{
In combination with flow-queuing, CoDel can reduce the experienced queuing delay, which is probably beneficial for Zoom's QoE, while equalizing the bandwidth share with TCP.
However, in low-bandwidth scenarios this share is still not perfectly equal.
Zoom does not reduce its rate but CoDel and FQ\_CoDel increasingly drop Zoom's packets which might affect Zoom's QoE negatively. A preliminary user study shows that FQ\_CoDel does, indeed, not improve QoE and can be found in the appendix.
}

\section{Conclusion}
\slabel{conclusion}
In this work, we recognize the impact of video conferencing on Internet stability and investigate congestion control fairness in combination with Zoom.
Flow-rate equality as fairness measure is well researched for TCP's congestion control and for real-world TCP flows in the Internet.
However, for congestion control of video conferencing software it is not -- specifically regarding different scenarios.
Hence, we investigate Zoom as increasingly popular real-world deployment of video conferencing.
We find that Zoom uses high shares of bandwidth in low-bandwidth scenarios yielding it when more bandwidth is available.
Adding AQM, such as CoDel, alone does not improve the bandwidth sharing, but reduces latency which is probably beneficial for Zoom.
Only when also using flow-queuing, more equal bandwidth sharing can be achieved with FQ\_CoDel.
However, this fair sharing comes at the price of reduced bandwidth and packet loss for Zoom, potentially reducing its QoE.
Our small-scale user study found that FQ\_CoDel did not improve the QoE.
For future work, we imagine a more thorough user study to evaluate Zoom's QoE with AQMs such as FQ\_CoDel in more detail.
Further, testing Zoom's reaction on ECN and multiple Zoom flows competing could give interesting information on its behavior on backbone congestion.

\section*{Acknowledgments}
This work has been funded by the Deutsche Forschungsgemeinschaft (DFG, German Research Foundation) under Germany's Excellence Strategy – EXC-2023 Internet of Production – 390621612.
We would like to thank the center for teaching- and learning services at RWTH Aachen University for issuing further Zoom licenses.
We further thank the anonymous reviewers and our shepherd Mirja Kühlewind for their valuable comments.
\bibliographystyle{splncs04}
\bibliography{reference}

\appendix
\section*{Appendix}
In the following, we present results of a small-scale user study which we conducted to analyze whether our findings regarding packet loss but also improvements regarding delay have positive or negative impact on Zoom's subjective quality.
However, as our study was performed with a limited number of participants due to COVID-19 restrictions, we had to restrict the number of scenarios that we could investigate.
Thus, the results and their generalizability are limited and this study should be regarded as an initial step in understanding how QoE, queuing and Zoom interact.

\section{QoE Impact of Flow-Queuing AQM}
\slabel{userstudy}

As we have shown in \sref{eval:fq_codel}, flow-queuing AQM can achieve more equal flow-rates and reduce latency when Zoom and TCP share a bottleneck.
However, this means lower bandwidths for Zoom, so likely worse video quality.
In contrast, lower latencies should probably mean better interactivity.
As the exact correlation w.r.t.\ perceived experience is hard to grasp, we perform a small-scale user study to capture the influence of flow-rate equality and AQM reduced latency on Zoom's QoE.

\afblock{Limitations of this Study.}
However, our study is limited, as we had to limit the number of participants (n=10) due to COVID-19 restrictions.
As such, we also restricted the number of scenarios to keep the individual study duration to roughly 25 minutes.
Additionally, we had to switch from synthetically generated videos (noise to maximize bandwidth utilization) that we used throughout \sref{eval} to real video-conferences.
This makes it difficult to compare the video-flows' demands from our synthetic evaluation to this user study as the bandwidth demand varies with the compression rate (higher compression for actual webcam video).
In summary, our study should only be regarded as an initial step.

In the following, we introduce the design and stimuli of our study and which metrics we are interested in.
Subsequently, we present the results.

\subsection{User Study Design}
We perform a video conference where the subject interacts with an experiment assistant via Zoom focusing on interactivity and understandability to rate the quality and whether potentially degraded quality is acceptable when a concurrent download is active.
The assistant reads short paragraphs of texts and the subject shortly summarizes them once the paragraph ended.
This way, we test whether the video conference allowed for easy understanding but also represent the typical condition where conference attendees interrupt each other unintentionally.
After 5 repetitions of summarizing, the subject and assistant alternately count to 10 to get a feeling for the delay, as proposed by the ITU~\cite{itut:p920}.
Lastly, the assistant reads random numbers and the subject stops the assistant at a given number (unknown to the assistant) for the same reasons.

\afblock{Quality Rating.}
After every run, the subject rates the overall, audio, video, and interactivity quality on a seven-point linear scale~\cite{itut:p851} (c.f., y-axis in \fref{eval:userstudy}).
Moreover, the subject decides (yes/no) if communicating was challenging, whether the connection was acceptable at all, whether the quality was acceptable if they were downloading a file during a business or private call or when someone else was downloading documents or watching movies in parallel.

\afblock{Test Conditions.}
We test 3 different scenarios using our previously described testbed; for all conditions, we shape the subject's link to \unit[0.5]{\mbps}, adjust the min.\ RTTs to \unit[50]{ms} and use a queue size of 10$\times$BDP.
The scenarios differ in whether an extra flow competes on the downlink and whether the queue is managed.
In detail, in Scenario~1 (Tail-Drop) only Zoom is active using a tail-drop queue.
Scenario~2 (Tail-Drop + Flow) adds a TCP CUBIC flow on the downlink, representing, e.g., a movie download.
Scenario~3 (FQ\_CoDel + Flow) adopts the TCP flow, but switches to the flow-queuing variant of CoDel.

\afblock{Study Details.}
We perform a ``within subject'' lab study: each subject rates every test condition selected from a latin square to randomize the order.
Each experiment takes about 5 min and is repeated for the 3 scenarios plus a training phase at the start using Scenario~1.
In total, the 4 experiments plus rating take about 25 min. %
Although conducting studies with members familiar to the study is discouraged~\cite{itut:p920}, we stick to the same experiment assistant to reduce variations.

\afblock{Subject Recruitment.}
Our subjects are 10 colleagues from our institute which volunteered to take part and are strongly familiar with Zoom.
We limited our study to these participants to reduce contacts during the pandemic.
As such we were able to hold the conferences in the participant's first language.

\subsection{Results}

\begin{figure}[t]
	\centering
	\includegraphics{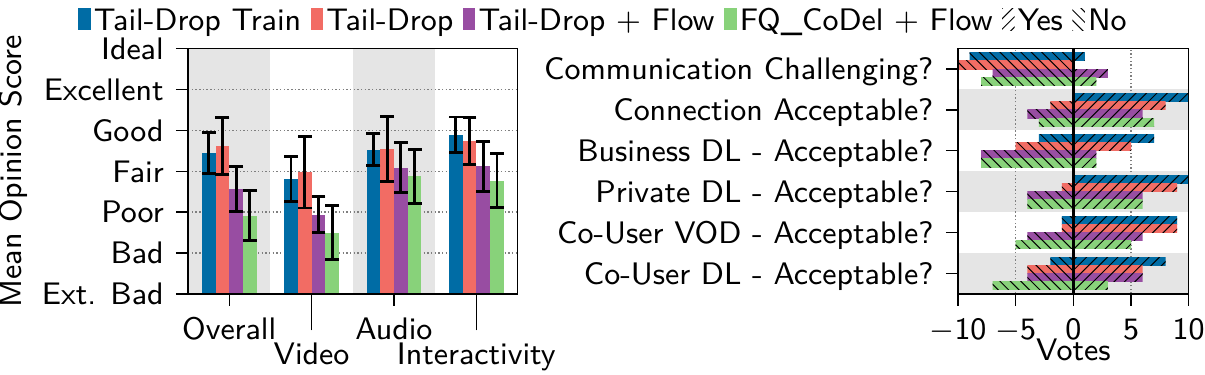}\vspace{-1em}
	\subfloat[\flabel{eval:userstudy:quality} Quality Rating]{\hspace{.57\linewidth}}
	\hspace{.19\linewidth}
	\subfloat[\flabel{eval:userstudy:votes} Acceptability]{\hspace{.23\linewidth}}
	\vspace{-0.5em}
	\caption{User Study Quality Rating and Votes}
	\flabel{eval:userstudy}
\end{figure}

\fref{eval:userstudy:quality} shows the mean opinion score and 95\% confidence intervals of the quality rating (distributions checked for normality via a Shapiro-Wilk test).
The confidence intervals are computed via the t-distribution due to our small sample size.
Further, \fref{eval:userstudy:votes} shows the distributions of ``Yes'' (positive, to the right) and ``No'' (negative, to the left) answers for the different questions.

Generally looking at the plots we can see that the worst results stem from using FQ\_CoDel, while using a tail-drop queue with no concurrent flow results in the best quality ratings.
For the overall quality of the video conference and the video quality this difference is statistically significant as the confidence intervals do not overlap.
However, for the scenarios where Zoom competes with TCP flows, the results are statistically insignificant and allow no statement.
Similar, all audio quality and interactivity votes allow no statistically significant statement.

\afblock{Flow-Queuing AQM induced QoE Changes.}
Hence, interpreting these results is complex.
What can be said is that CoDel's positive effect of reducing the queuing delay was not perceived by the users.
On the other hand, also the reduction in bandwidth did not yield any statistically significant quality reduction.
However, a trend against using FQ\_CoDel is visible, but it cannot be statistically reasoned.
Only following the trend, it might be not worth using  FQ\_CoDel due to its potentially worse QoE.
Otherwise, only few users considered the connection unacceptable (c.f.\ \fref{eval:userstudy:votes}), surprisingly uncorrelated to whether FQ\_CoDel was used or whether a concurrent flow was actually started.
I.e., some users considered our scenarios generally as unacceptable regardless of FQ\_CoDel.

\afblock{Influence of Concurrent Downloads on Acceptability.}
Surprisingly, users also consider the quality unacceptable when imagining a concurrent download of documents in business or private conversations.
We expected that users accept deteriorations, as they would not pay attention to the video conference, but want their download to complete.
However, specifically in the business case, our users did not.
Also quality deteriorations induced by other users downloading movies or documents were not seen more disturbing.
I.e., independent of self-inflicted or not, some users do not accept quality deteriorations at all, while others do.

\takeaway{
Unfortunately, our study did not yield statistically conclusive results with respect to how participants perceive the difference in Zoom quality between using a tail-drop queue and FQ\_CoDel when a flow competes.
Also regarding acceptance, users did not see strong differences and either disliked the quality regardless of possible concurrent downloads as reasons or just accepted it, disagreeing on a generally applicable statement.
Looking at the general trend of our study, FQ\_CoDel could decrease QoE.
}

\end{document}